\documentclass
[floats,prb,noshowpacs,balancelastpage,10pt,byrevtex,letterpaper,preprint,twocolumn]{revtex4}%
\usepackage{amsfonts}
\usepackage{amsmath}
\usepackage{amssymb}
\usepackage{graphicx}%
\begin{document}
\title{Manifestations of the absence of spin diffusion in multipulse NMR\ experiments
on diluted dipolar solids}
\author{Mar\'{\i}a Bel\'{e}n Franzoni}
\affiliation{Facultad de Matem\'{a}tica, Astronom{\'{\i}}a y F{\'{\i}}sica, Universidad
Nacional de C\'{o}rdoba, Ciudad Universitaria, 5000, C\'{o}rdoba, Argentina}
\author{Patricia R. Levstein}
\affiliation{Facultad de Matem\'{a}tica, Astronom{\'{\i}}a y F{\'{\i}}sica, Universidad
Nacional de C\'{o}rdoba, Ciudad Universitaria, 5000, C\'{o}rdoba, Argentina}

\begin{abstract}
Puzzling anomalies previously observed in multipulse NMR experiments\ in
natural abundance $^{29}$Si [A.E. Dementyev, D. Li, K. MacLean, and S.E.
Barrett, Phys. Rev. B \textbf{68}, 153302 (2003)] such as long-lived spin
echoes and even-odd asymmetries, are also found in polycrystalline C$_{60}$.
Further experiments controlling the phases and tilting angles of the pulse
trains, as well as analytical and numerical calculations allowed us to explain
\ the origin of these anomalies. We prove that the observation of long
magnetization tails requires two conditions: i) an rf field inhomogeneity able
to produce different tilting angles in different sites of the sample and ii)
the absence of spin diffusion (non-effective flip-flop interactions). The last
requirement is easily satisfied in diluted dipolar solids, where the frequency
differences between sites, caused by disorder or other sources, are usually at
least one order of magnitude larger than the dipolar couplings. Both
conditions lead to the generation of stimulated echoes in Carr-Purcell (CP)
and Carr-Purcell-Meiboom-Gill (CPMG)\ pulse trains. We show, both
experimentally and theoretically, that the stimulated echoes interfere
constructively or destructively with the normal (Hahn) echoes depending on the
alternation or not of the $\pi$ pulse phases in the CP and the CPMG sequences.
Constructive interferences occur for the CP and CPMG sequences with and
without phase alternation respectively, which are the cases where long
magnetization tails are observed. Sequences with two, three and four $\pi$
pulses after the $\pi/2$ pulse allow us to disentangle the contributions of
the different echoes and show how the stimulated echoes originate the even-odd
asymmetry observed in both $^{29}$Si and C$_{60}$ polycrystalline samples.

\end{abstract}
\startpage{1}
\maketitle

During the last years, the control of quantum coherence has become of great
importance for quantum information processing and for the development of new
nanotechnologies. In the nanoscale, many quantum phenomena become evident and
can be exploited. However, interaction with various degrees of freedom of the
environment degrades the quantum coherence. As a consequence, it is necessary
to understand in detail the mechanisms that produce decoherence and their
timescales. Many suggestions to implement quantum computations involve nuclear
or electron spins.\cite{QCompgral,Cory97} In particular, as silicon has been
an essential component of one of the main proposals,\cite{Kane} a group of
researchers carried out a series of nuclear magnetic resonance (NMR)
measurements to study the spin-spin \ "decoherence" time ($T_{2}$) in $^{29}%
$Si ($4.67\%$ natural abundance, spin $\frac{1}{2}$)\ doped with different
concentrations of donors or acceptors.\cite{barrett} Their results were
completely unexpected and defied explanation to date.

They performed $T_{2}$ measurements with different pulse sequences commonly
used with this purpose. They found that it is possible to detect \ $^{29}$Si
NMR signals up to much longer times (more than two orders of magnitude) by
applying the Carr-Purcell-Meiboom-Gill sequence as compared with the
characteristic decay time obtained from the Hahn echo sequence ( $\approx$ 5.6
ms). As these sequences usually give similar results, the difference cannot be
disregarded and requires an explanation.

Several tests were performed to determine the origin of this anomalous
behavior. First of all, it was evident that there was no dependence on the
amount of donors or acceptors in the $^{29}$Si sample, even when this could
change the linewidth and the spin-lattice relaxation time in one order of
magnitude. Moreover, they found the same behavior in a pure single crystal of
n.a. $^{29}$Si.\cite{APScommunication}

One hint was that the anomaly seemed to be present in systems with diluted
spins. Thus, we implemented a series of related experiments and calculations
in a C$_{60}$ sample. As each molecule has sixty carbon atoms and the magnetic
isotope $^{13}$C, has a natural abundance $1,1\%$, in a typical sample there
are $34.5\%$ of molecules containing one $^{13}$C, $11.4\%$ containing exactly
two, and $2.4\%\ $containing three.

As a first step, we search for an approximation to $T_{2}$ using the Hahn echo
sequence\cite{Hahn} $[HE:\left(  \frac{\pi}{2}\right)  _{X}-\tau-\left(
\pi\right)  _{Y}-\tau-echo],$ where $\tau$ is a variable time. We obtained
$T_{2}^{HE}\simeq15\ $ms. Then, we used different train pulse sequences to
measure $T_{2}$, derived from the Carr-Purcell sequence,\cite{Carr_Purcell}
\begin{align}
CP1 &  :\left(  \frac{\pi}{2}\right)  _{X}-\left[  \tau-\left(  \pi\right)
_{X}-\tau-echo\right]  _{n};\label{CP}\\
CP2 &  :\left(  \frac{\pi}{2}\right)  _{X}-\left[
\begin{array}
[c]{c}%
\tau-\left(  \pi\right)  _{X}-\tau\ -echo-\tau-\\
-\left(  \pi\right)  _{-X}-\tau-echo
\end{array}
\right]  _{n},\nonumber
\end{align}
and the Carr-Purcell-Meiboom-Gill sequence,\cite{fuku}
\begin{align}
CPMG1 &  :\left(  \frac{\pi}{2}\right)  _{X}-\left[  \tau-\left(  \pi\right)
_{Y}-\tau-echo\right]  _{n};\label{CPMG}\\
CPMG2 &  :\left(  \frac{\pi}{2}\right)  _{X}-\left[
\begin{array}
[c]{c}%
\tau-\left(  \pi\right)  _{Y}-\tau\ -echo-\tau-\\
-\left(  \pi\right)  _{-Y}-\tau-echo
\end{array}
\right]  _{n}.\nonumber
\end{align}

Here, the numbers $1$ and $2$ have been added to the names of the sequences to
denote one or two different phases for the $\pi$ pulses.

We found that by applying the $CPMG1$ sequence or the $CP2$ sequence, a
magnetization tail appears, i.e. the signal remains for times much longer than
$T_{2}^{HE}.$ Moreover, with both of them we found that by choosing $2\tau$
$\geqslant T_{2}^{HE}$ the magnetization shows an $\emph{even-odd\ behavior}$
as observed in $^{29}$Si, \cite{barrett} (that is when odd echoes are already
in the noise, even echoes are still visible).

Another unexpected observation was the presence of noticeable \emph{stimulated
echoes} when applying pulse sequences of the form, $\ \left[  SE:\left(
\frac{\pi}{2}\right)  _{X}-\tau-\left(  \pi\right)  _{\varphi_{2}}%
-t_{1}\ -\left(  \pi\right)  _{\varphi_{3}}-acq\right]  $ with variable $\tau$
and $t_{1}$. In principle, these sequences involving $\pi$ pulses should not
produce stimulated echoes. As we will show, the constructive or destructive
interferences between the stimulated and\ \emph{normal }echoes enable a
phenomenological explanation for the long magnetization tails.

\begin{figure}
[ptbh]
\begin{center}
\includegraphics[
width=9cm
]%
{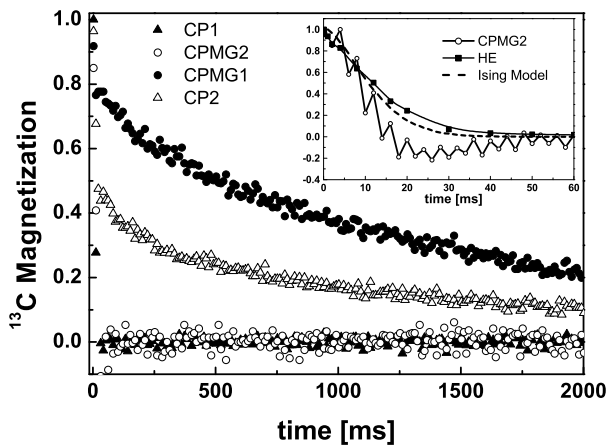}%
\caption{Comparison of the $^{13}$C magnetizations acquired with different
$T_{2}\ $sequences ($\tau=1$ ms). The insert shows the decay observed with the
Hahn echo, the decay simulated by considering only the Ising terms of the
dipolar interaction and the short time regime of the $CPMG2$ sequence.}%
\label{secuencias}%
\end{center}
\end{figure}

Figure \ref{secuencias} shows the normalized magnetizations acquired by
applying the sequences $HE$; $CP1$; $CP2$; $CPMG1$ and $CPMG2$ to
polycrystalline C$_{60}$. In the last four sequences $\tau=1$ms and signal
acquisition is performed at the top of the echo. In all the sequences the
durations of the $\pi$ and $\pi/2$\ pulses\ were carefully set from nutation
experiments to 7.0 $\mu s$ and 3.4 $\mu s,$ respectively. No effort was done
to reduce the rf field inhomogeneity which, for the completely full sample
holder, was in the range 15-20\%. It is clear from Fig.(\ref{secuencias}) that
the magnetization, measured with the $CP2$ and $CPMG1$ sequences, lasts for
very long times ($\approx$1s).

On the other hand, by applying the $CP1$ or the $CPMG2$ sequences the decay
time is not longer than $T_{2}^{HE}$, but it is remarkable that at short times
the magnetization makes a zigzag that, as will become clear, it is also a
consequence of the presence of stimulated echoes. Besides, it should be
noticed that the magnetization goes through negative values before reaching
its final asymptotic zero value.%

The first question was if the $T_{2}^{HE}$ measured for the C$_{60}$ sample
could be directly assigned to the dipolar interaction within the $^{13}$C
nuclei. Our measurements of the spin-lattice relaxation time yielded
$T_{1}\simeq35$ s reflecting that there are no efficient mechanisms for this
kind of relaxation that should be taken into account in our analysis of the
\ "decoherence" time. Thus, the main interactions in this system led us to
write the Hamiltonian in the rotating frame as%

\begin{equation}
\mathcal{H}=\sum_{i=1}^{N}\left[  \hbar\delta\omega_{i}I_{i}^{z}+\sum
_{j>i}^{N}\left(  a_{ij}I_{i}^{z}I_{j}^{z}+b_{ij}\left(  I_{i}^{+}I_{j}%
^{-}+I_{i}^{-}I_{j}^{+}\right)  \right)  \right]  \text{,}
\label{hamiltoniano}%
\end{equation}
where $\delta\omega_{i}$ represents the chemical shift for spin $i$ (relative
to on-resonance spins); $a_{ij}=(\gamma^{2}\hbar^{2}/r_{ij}^{3})(1-3\cos
^{2}\theta_{ij})$; $b_{ij}=-a_{ij}/4$ with $\theta_{ij}$ the angle between the
internuclear vector $\mathbf{r}_{ij}^{{}}$ and the static magnetic field. The
gyromagnetic ratio of $^{13}$C is $\gamma=6.7283\times10^{7}$rad T$^{-1}%
$s$^{-1}$ and the center to center distance between neighboring molecules is
$r_{ij}=10$ \AA .\cite{C60} It should be noticed that because of the rapid
isotropic rotations of the buckyballs in the solid, the spins can be
considered to lie at the center of the molecule.\cite{Tycko}

The Lorentzian full width at half maximum, $FWHM\simeq167$ Hz, corresponding
to a free induction decay time $T_{2}^{\ast}\simeq1.91$ ms is almost an order
of magnitude shorter than $T_{2}^{HE}\simeq15.0\ $ms, thus reflecting the
sample inhomogeneity. Then, it is usual to make the unlike spins approximation
$\left\vert a_{ij}\right\vert \ll\left\vert \delta\omega_{i}-\delta\omega
_{j}\right\vert $, and truncate the $b_{ij}$ terms, leaving only the Ising
contribution. Under these conditions, one can use the product operator
formalism, \cite{Slichter} to obtain an analytical expression for the
magnetization. Starting from the initial equilibrium density operator in the
conventional strong field and high temperature approximations:%

\begin{equation}
\rho(t=0)\propto\sum_{i}I_{i}^{z}\text{,} \label{rhoinicial}%
\end{equation}
the application of the Hahn echo sequence with variable inter-pulse time
$\tau=k\tau,$ leads to the expression
\begin{equation}
\left\langle I_{y}(k2\tau)\right\rangle =\sum_{i}^{N}I_{_{i}}^{y}(0)\left\{
\prod_{j>i}^{N}\cos(a_{ij}k\tau)\right\}  \text{.} \label{Iy}%
\end{equation}

Our calculation for a polycrystalline sample of C$_{60}$ involves a set of 600
pieces ($\approx$ 4000 unit cells each) with different representative
orientations with respect to the external magnetic field and a random
distribution of molecules with one, two or three spins each considering the
proportions expressed above. The characteristic decay time of the analytic
curve is in very good agreement with the experimental one obtained from the
Hahn echo sequence (see Ising model in Fig \ref{secuencias}), manifesting that
$T_{2}^{HE}$ can be directly associated with the dipolar interaction.

\begin{figure}
[ptb]
\begin{center}
\includegraphics[
width=9cm
]%
{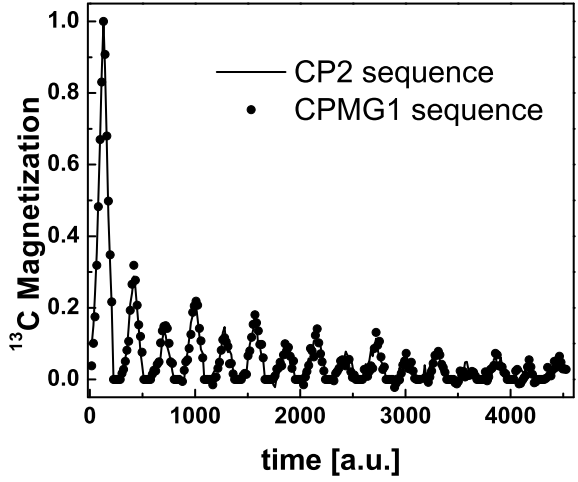}%
\caption{Even-odd effect observed with the $CPMG1$ and $CP2$ sequences in a
C$_{60}$ polycrystalline sample. The interpulse separation is $\tau=8$ ms.}%
\label{parimpar}%
\end{center}
\end{figure}

In order to look for the even-odd behavior reported for the Si
sample\cite{barrett} in C$_{60}$, we took long inter-pulse separations
$(2\tau>T_{2}^{HE})$. We observed the anomalous behavior by applying the
$CPMG1$ and the $CP2$ sequences, as displayed in Fig.\ref{parimpar} for
$\tau=8$ ms. While the first echo, occurring at $t=16$ ms is higher than the
second one, occurring at $t=32$ ms, the third is lower than the forth, and so
on. Moreover, the same sequence with $\tau=25$ ms produces a first echo which
is smaller than the second one.%

So far, all the results reported in silicon were also manifested in the
C$_{60}$ sample, but there was still one more thing to check, the presence of
stimulated echoes.

We applied a sequence of three pulses \cite{Hahn}%
\begin{equation}
SE:\ \left(  \frac{\pi}{2}\right)  _{\varphi_{1}}-\tau-\left(  \pi\right)
_{\varphi_{2}}-t_{1}\ -\left(  \pi\right)  _{\varphi_{3}}-acq,\label{se}%
\end{equation}
taking $\varphi_{1}=X$ in all the experiments and varying $\varphi_{2}$ and
$\varphi_{3}$ in correspondence with the sequences we used in the first
experiments. For these stimulated echo sequences, we denote $\left[
CPMG1^{SE}:\varphi_{2}=Y;\varphi_{3}=Y\right]  $, $\left[  CP1^{SE}%
:\varphi_{2}=X;\varphi_{3}=X\right]  $, $\left[  CPMG2^{SE}:\varphi
_{2}=Y;\varphi_{3}=-Y\right]  $, and $\left[  CP2^{SE}:\varphi_{2}%
=X;\varphi_{3}=-X\right]  .$ \ At a time $t_{1}-\tau$ after the third pulse,
the normal echo (i.e. the refocusing of the Hahn echo at time $\tau$ after the
second pulse) appeared, and for all the sequences a stimulated echo peaked at
time $\tau$ after the third pulse.

\begin{figure}
[ptb]
\begin{center}
\includegraphics[
width=9cm
]%
{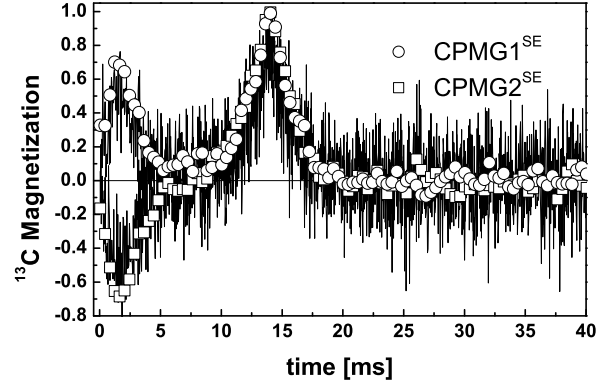}%
\caption{Stimulated echoes observed with the $CPMG1^{SE}$ and $CPMG2^{SE}$
sequences in a C$_{60}$ polycrystalline sample. The parameters are $\tau=1$ ms
and $t_{1}=15$ ms, giving rise to a stimulated and a normal echo at $1$ ms and
$14$ ms after the last pulse, respectively. The circles and squares represent
splines (average of 20 original datapoints per symbol) performed to guide the
eye. Notice the inversion of the phase in the stimulated echo.}%
\label{estimulados}%
\end{center}
\end{figure}

A very remarkable fact is that by applying the $CPMG1^{SE}$ or the $CP2^{SE}$
sequence, the stimulated and normal echoes have the same phases, but when
applying the $CPMG2^{SE}$ or the $CP1^{SE}$ sequences they have opposite
phases, as shown in Fig.\ref{estimulados} for the $CPMG$ variations.%

Using the superoperator formalism, \cite{Mehring,Ernst} for an arbitrary
three-pulse sequence $\left[  \left(  \beta_{1}\right)  _{\varphi_{1}}%
-\tau-\left(  \beta_{2}\right)  _{\varphi_{2}}-t_{1}-\left(  \beta_{3}\right)
_{\varphi_{3}}-acq\right]  \ $where $\beta_{i}$ and $\varphi_{i}$ are
arbitrary tilting angles and phases in the $XY$ plane respectively, we
analytically solved the dynamical equations in the delta-pulse approximation.
By taking into account that the $FWHM\simeq167$ Hz and $(\pi T_{2}^{HE}%
)^{-1}\simeq21.2$ Hz, we can neglect the spin-spin interactions and work with
the Hamiltonian $\mathcal{H}=\hbar\sum_{i=1}^{N}\delta\omega_{i}I_{i}^{z},$\ 

and the initial condition given in Eq.(\ref{rhoinicial}). Under these
approximations, the following response function is obtained%

\begin{eqnarray}
G(t >\tau+t_{1})=&e^{i\varphi_{1}}f_{1}G(t)&+e^{i\left(  2\varphi
_{3}-2\varphi_{2}+\varphi_{1}\right)  }f_{2}G(t-2t_{1})\nonumber\\
+& e^{i\left(  \pi+2\varphi_{2}-\varphi_{1}\right)}&f_{3}G(t-2\tau
)\nonumber\\+&e^{i(\pi+2\varphi_{3}-\varphi_{1})}&f_{4}G(t-2t_{1}-2\tau)\nonumber\\
  +&e^{i(\pi+\varphi_{3}-\varphi_{2}+\varphi_{1})}&f_{5}G\left(  t-t_{1}%
\right)\nonumber\\ +&e^{i(\pi+\varphi_{3}+\varphi_{2}-\varphi_{1})}&f_{6}G(t-t_{1}%
-2\tau)\nonumber\\
+&e^{i(\pi+2\varphi_{3}-\varphi_{2})}&f_{7}G(t-2t_{1}-\tau),\label{GT}%
\end{eqnarray}
where $G(t)=\frac{1}{N}\sum_{j=1}^{N}\exp(it\delta\omega_{j})$ represents the
normalized FID, the time $t$ is the total time starting at the very first
pulse, and the signal amplitudes $f_{j}\equiv f_{j}(\beta_{1},\beta_{2}%
,\beta_{3})$ are determined from the exact calculation:%

\begin{align*}
f_{1}  &  =\sin\beta_{1}\cos^{2}\frac{\beta_{2}}{2}\cos^{2}\frac{\beta_{3}}%
{2}\\
f_{2}  &  =\sin\beta_{1}\sin^{2}\frac{\beta_{2}}{2}\sin^{2}\frac{\beta_{3}}%
{2}\\
f_{3}  &  =\sin\beta_{1}\cos^{2}\frac{\beta_{3}}{2}\sin^{2}\frac{\beta_{2}}%
{2}\\
f_{4}  &  =\sin\beta_{1}\cos^{2}\frac{\beta_{2}}{2}\sin^{2}\frac{\beta_{3}}%
{2}\\
f_{5}  &  =\frac{1}{2}\sin\beta_{1}\sin\beta_{2}\sin\beta_{3}\\
f_{6}  &  =\frac{1}{2}\sin\beta_{1}\sin\beta_{2}\sin\beta_{3}\\
f_{7}  &  =\cos\beta_{1}\sin\beta_{2}\sin^{2}\frac{\beta_{3}}{2}%
\end{align*}

The stimulated echo signal is the one that peaks at total time $t=t_{1}+2\tau
$. Its amplitude is given by the expression $f_{6},$ evidencing that for
sequences like the ones applied here (i.e. $\beta_{2}=\beta_{3}=\pi$), it
cancels out. Consequently, the stimulated echo should not appear.

The \emph{normal echo},\emph{ }second term in Eq.(\ref{GT}), peaks at a total
time $t=2t_{1}$. Its phase in the $CPMG1$ ($\varphi_{1}=0$, $\varphi_{2}%
=\pi/2,$ $\ \varphi_{3}=\pi/2$) or in the $CP2$ ($\varphi_{1}=0$, $\varphi
_{2}=0,$ $\ \varphi_{3}=\pi$) sequence is $e^{i\left(  2\varphi_{3}%
-2\varphi_{2}+\varphi_{1}\right)  }=1$ coinciding with that of the
\emph{stimulated echo} $e^{i(\pi+\varphi_{3}+\varphi_{2}-\varphi_{1})}=1$.
However, if we apply the $CPMG2$ ($\varphi_{1}=0$, $\varphi_{2}=\pi/2,$
$\ \varphi_{3}=-\pi/2$) or the $CP1$ ($\varphi_{1}=0$, $\varphi_{2}=0,$
$\ \varphi_{3}=0$) sequences they are in opposite phases, because the phase of
the stimulated echo is inverted. Here, we have used $\varphi_{i}=0$ for the
$X$ axis and it should be noticed that under the experimental conditions
($\beta_{2}=\beta_{3}$) $f_{2}$ and $f_{6}$ are always positive. Then, the
remarkable fact is that these analytical calculations agree with the
experimental result for any pair of tilting angles $\beta_{2}=\beta_{3}\neq
\pi$. This lead us to conclude that not all the spins are affected by $\pi$
pulses. We performed the three-pulse experiment for $\varphi_{1}=0$,
$\varphi_{2}=\pi/2,$ $\ \varphi_{3}=\pi/2$, $\beta_{1}=\pi/2$\ and\ $\beta
_{2}=\beta_{3}$ varying in the range $\left[  0,5\pi/2\right]  $ with
$\tau=14$ ms and $t_{1}=15$ ms. While the experiments reproduce the functional
form of $f_{6}$, there are no zeroes in the magnitude of the stimulated echoes
for any value of $\beta_{2}$. It is noticeable that the minimum magnitude of
the SE, obtained for $\beta_{2}=\beta_{3}\approx\pi$ is approximately 30\% of
the maximum value that occurs for $\beta_{2}=\beta_{3}=\pi/2$ (data not shown).

It is evident that if we take $t_{1}=2\tau$ in the three-pulse sequences
(\ref{se}) the normal and the stimulated echoes will appear at the same time,
and as can be seeing in fig (\ref{estimulados}), they will interfere
constructively in the $CPMG1$ and the $CP2$ sequences or destructively in the
$CPMG2$ and the $CP1$ ones. As a consequence of these constructive
interferences and the well known fact that stimulated echoes have much longer
decay times,\cite{Hahn,Mehring} the magnetization tails appear.

\begin{figure}
[ptb]
\begin{center}
\includegraphics[
width=9cm
]%
{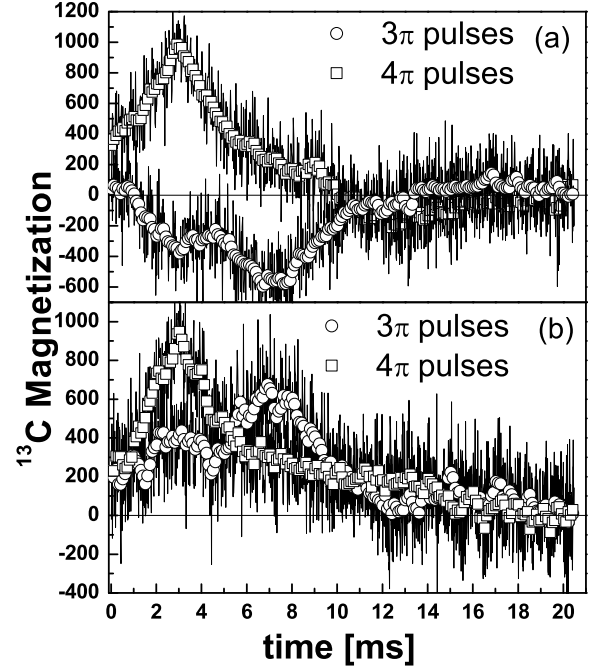}%
\caption{Echoes obtained after three and four $\pi$ pulses with the (a)
$CP2^{SE}$ and (b) $CPMG1^{SE}$-like sequences. The time between the $\pi/2$
and the $\pi$ pulse is $\tau=3$ ms, while the intervals between $\pi$ pulses
are $t_{1}=10$ ms. For three $\pi$ pulses the stimulated echo appears at
$t_{1}-\tau=7$ ms after the last pulse, while the normal echo appears at
$\tau.$ The normal \ and stimulated echo times after the last $\pi$ pulse are
exchanged in the case of four $\pi$ pulses. Notice that after four $\pi$
pulses, the normal echoes have almost dissapeared. The circles and squares
represent splines (20 original datapoints averaged).}%
\label{3y4pi}%
\end{center}
\end{figure}

In order to analyze the different contributions to the even-odd asymmetry in
the echoes, displayed in Fig.\ref{parimpar}, we performed experiments with
two, three and four $\pi$ pulses after the $\pi/2$ pulse with interpulse
separations precluding the overlap of the normal and stimulated echoes. In
particular, we set the separation between the $\pi/2$ and the $\pi$ pulses
$\tau=3$ ms, while the separations between $\pi$ pulses where $t_{1}=10$ ms.
The results for three and four $\pi$ pulses for the case of a $CP2^{SE}$ and
$CPMG1^{SE}$-like sequences are shown in Fig. \ref{3y4pi} (a) and (b)
respectively.

As can be seen, the stimulated echo occurring after four $\pi$ pulses, i.e. at
total time $t=3t_{1}+2\tau$ or $\tau$ after the last pulse, is larger than the
one after three $\pi$ pulses $t=3t_{1}$(or\ $t_{1}-\tau$ after the last pulse)
in both cases. We use the same formalism and Hamiltonian applied for the
three-pulse sequences, setting a single tilting angle $\beta$ for all the
pulses after the $\pi/2$ one, with $\varphi_{1}=0$ and $\varphi_{i}=0$ or
$\pi$ alternatively for $i\neq1$ ($CP2^{SE}$) or $\varphi_{1}=0$ and
$\varphi_{i}=\pi/2$ for $i\neq1$( $CPMG1^{SE}$). Thus, we obtained for the
stimulated echo amplitudes the functions: $A_{4\pi}=1/2\sin^{2}\beta
(5/2\cos^{2}\beta-\cos\beta+1/2)$ and $A_{3\pi}=\mp1/2\sin^{2}\beta
(1-\cos\beta)$. The minus sign in $A_{3\pi}$\ corresponds to the $CP2^{SE}$
sequence. Again, it is evident that for $\beta=\pi,$ there should be no
stimulated echo, i.e. $A_{4\pi}=A_{3\pi}=0.$ However, it can be seen from
these formulae that for angles in the interval $\beta=\pi(1\pm0.35)$, one gets
$A_{4\pi}>A_{3\pi},$ which is in qualitative agreement with the experimental
observation. Numerical calculations for longer sequences (i.e. a larger number
of $\pi$\ pulses) support this alternation of amplitudes. Consequently, having
proved that the stimulated echoes interfere constructively with the normal
echoes in the sequences $CPMG1$ and $CP2$, we have found the explanation for
the even-odd asymmetry observed in Fig.\ref{parimpar}.

In order to further verify the fact that not all the sites in the sample are
experiencing the same externally fixed tilting angle, we applied the same
sequences (\ref{CP}) and (\ref{CPMG}) but applying our best $2\pi$ or $4\pi$
pulses (zeroes in the nutation experiment), instead of $\pi$ pulses. In all of
them, we kept the first $\left(  \frac{\pi}{2}\right)  _{X}$ pulse. Under
these conditions, \ if the pulses were perfect, one would expect to acquire
just an FID signal. This is exactly what we got by applying the $CPMG2_{(2\pi
)}$ or the $CP1_{(2\pi)}$ sequences or their equivalent with $4\pi$ pulses.
Instead, by applying the $CPMG1_{(2\pi)}$ or the $CP2_{(2\pi)}$ sequences the
magnetization tails reappeared as also occurred with the $4\pi$ pulses.

The conclusion that not all the spins experience the same tilting pulse should
not be surprising, as it is well known that the rf field inhomogeneity may be
around 5-10\% in ordinary experimental conditions. However, the remaining
question is why these long tails do not appear when working with magnetically
concentrated samples,\cite{Ladd05} as usually occurs in solid-state $^{1}%
$H-NMR measurements. While the rf field inhomogeneities occur in all cases, in
diluted systems, the dynamical contribution of the dipolar interaction (i.e.
the flip-flop part) is not effective. Then, it can be conceived that in
concentrated systems, there are different tilting angles in different sites,
but during the interpulse time the defects or excesses in the $\pi$ pulses are
exchanged among the sites because of the flip-flop, thus, effectively
averaging to the externally fixed pulse. By contrast, in diluted systems the
imperfection over one site has always the same sign or direction, originating
the stimulated echoes and consequently the long tails. This fact was verified
through numerical simulations where imperfections up to 10\% in the $\ $"$\pi$
pulses" were randomly applied to different spin sites. But a very important
ingredient in this explanation is that the \emph{disorder in energies} among
the different sites must be greater than the dipolar couplings. This condition
occurs quite naturally in diluted systems because the dipolar interaction is
very small; thus, any interaction capable to produce differences in site
energies will be enough to hinder the flip-flop dynamics. As a practical
warning, one should consider the fact that magnetically diluted samples
require larger amounts making more serious the problem of rf field inhomogeneity.

The long tails obtained in our experiments can not be interpreted as a
decrease in the decoherence rate of the spin system. Because of the pulse
imperfections, some magnetization is preserved in the direction of the
external magnetic field, having a decay time in the order of the spin-lattice
relaxation time.\cite{Mehring}

Our findings may be significant for qubit coherence control in a nuclear spin
bath, as occurs with electron spins in a lattice containing Si, GaAs, In or
Al.\cite{Sousa05,Abe04} In these situations, the electron spin coherence is
limited by spectral diffusion arising from dipolar fluctuations of lattice
nuclear spins.

Another important aspect to discuss is that, as it is well known, the dipolar
dynamics is reversible,\cite{Rhim,ZME92,LUP98} and there is no real loss of
information in the collective spin system because of it. In this sense, a more
relevant measurement of \ "decoherence time" in C$_{60}$, would involve an
homonuclear decoupling as performed by Ladd et al.\cite{Ladd05} in silicon.
Under those conditions, more detailed experiments are required to test if
there is still a contribution from stimulated echoes. We find no reason to
discard it beforehand, because the necessary conditions: rf inhomogeneity and
no flip-flop dynamics will still be present.

The absence of nuclear spin diffusion observed in our experiments and in those
performed in n.a. $^{29}$Si,\cite{barrett} seems to be a particular
manifestation of the \emph{localization} phenomenon.\cite{Anderson2, Anderson}
However, as one is working with a 3-d many-body system, the complexity of a
theoretical approach is beyond the scope of this work. Nevertheless,
NMR\ studies as a function of nuclear spin concentrations and using different
time windows may help to search for a possible phase transition, opening up a
new way to look at localization phenomena.\bigskip

We acknowledge support from Fundaci\'{o}n Antorchas, CONICET, FoNCyT, and
SeCyT-UNC. P.R.L. is a member of the Research Career and M.B.F a doctoral
fellow of CONICET. We benefitted from fruitful discussions with H.M. Pastawski
on localization phenomena.

\bigskip

\bibliographystyle{apsrev}
\bibliography{acompat,referencias}

\end{document}